\documentclass{IEEEtran}

\usepackage{graphics,graphicx}
\usepackage{epsfig}
\usepackage{amsmath}
\usepackage{amssymb}
\usepackage{tikz}
\usepackage{amsthm}
\usepackage{comment}
\usepackage[export]{adjustbox}
\usepackage{mathcomSTEP}
\usepackage{subcaption,setspace}
\usepackage[utf8]{inputenc}

\usepackage{tikz}
\usetikzlibrary{positioning}
\allowdisplaybreaks

\tikzstyle{int}=[draw, fill=blue!10, minimum height = 1cm, minimum width=1.5cm,thick ]
\tikzstyle{joint} = [draw, circle, minimum size=1em]

\theoremstyle{remark}

\newtheorem{definition}{Definition}

\usepackage{tikz}
\usetikzlibrary{positioning}
\usetikzlibrary{arrows,fit}

\usepackage{epstopdf}
\usepackage{pgfplots,setspace}

\tikzstyle{int}=[draw, fill=blue!10, minimum height = 1cm, minimum width=1.5cm,thick ]
\tikzstyle{sum}=[circle, fill=black!10, draw=black,line width=1pt,minimum size = 0.3cm, thick ]
\tikzset{cross/.style={cross out, draw=black, minimum size=2*(#1-\pgflinewidth), inner sep=0pt, outer sep=0pt},
cross/.default={1pt}
}
\usetikzlibrary{shapes.misc}

\begin{document}

\title{
The Carbon Copy onto Dirty Paper Channel \\
with Statistically Equivalent States
}

\author{%
\IEEEauthorblockN{%
Stefano Rini \IEEEauthorrefmark{1} and Shlomo Shamai  (Shitz)  \IEEEauthorrefmark{2} \\}
\IEEEauthorblockA{%
\IEEEauthorrefmark{1}
National Chiao-Tung University, Hsinchu, Taiwan\\
E-mail: \texttt{stefano@nctu.edu.tw} }

\IEEEauthorblockA{%
\IEEEauthorrefmark{2}
Technion-Israel Institute of Technology,  Haifa, Israel \\
E-mail: \texttt{sshlomo@ee.technion.ac.il}
}

%
}
\maketitle

\begin{abstract}
Costa's ``writing on dirty paper'' capacity result establishes that full state pre-cancellation can be attained in Gel'fand-Pinsker channel
with additive state and additive Gaussian noise.
The ``carbon copy onto dirty paper'' channel is the extension of Costa's model to the compound setting:
$M$ receivers each observe the sum of the channel input, Gaussian  noise and one of $M$ Gaussian state sequences
and attempt to decode the same common message.
The state sequences are all non-causally known at the transmitter which  attempts to simultaneously pre-code its transmission against the channel state affecting each output.
In this correspondence we derive the capacity to within $2.25$ bits-per-channel-use of the carbon copying onto dirty paper channel in which the state sequences
are statistically equivalent, having the same variance and the same pairwise correlation.
For this channel capacity is approached by letting the channel input be the superposition of two codewords: a base codeword, simultaneously decoded at each user,
and a top codeword which is pre-coded against the state realization at each user for a portion $1/M$ of the time.
The outer bound relies on a recursive bounding in which incremental side information is provided at each receiver.
%
This result represents a significant first step toward determining the capacity of the most general ``carbon copy onto dirty paper'' channel in which state sequences appearing in
the different channel outputs have any jointly Gaussian distribution.
\end{abstract}

\begin{IEEEkeywords}
Gel'fand-Pinsker Problem;
Compound State-Dependent Channel;
Carbon Copying onto Dirty Paper;
\end{IEEEkeywords}
In the Gel'fand-Pinsker (GP) channel \cite{GelfandPinskerClassic} the output of a point-to-point channel is obtained as a random function of the channel input
and a state sequence which is provided non-causally to the encoder but is unknown at the decoder.
Costa`s ``Writing on Dirty Paper'' (WDP) channel \cite{costa1983writing} is the Gaussian version of the GP channel in which the channel output is obtained as a linear
combination of the input, the state sequence and iid, Gaussian-distributed, noise.
Perhaps surprisingly, Costa showed that the capacity of the WDP channel is the same as the capacity of the point-to-point channel
in which the state is not present in the channel output.
In other words, the transmitter can fully pre-code its transmissions against the  channel state and thus the presence of the channel state does not affect capacity.
The ``Carbon Copying onto Dirty Paper'' (CCDP) channel \cite{LapidothCarbonCopying} is the extension of the GP channel to the compound scenario:
in this model  the transmitter wishes to communicate the same message to $M$ receivers which observe as channel output the summation of the channel input,
iid Gaussian noise and one of $M$ state sequences.
These state sequences are all provided non-causally to the transmitter but are unknown at the receivers.

In this correspondence we derive the capacity of the CCDP channel with any number of users for the case in which the states are statically equivalent, being Gaussian-distributed
with the same variance and the same pairwise correlation.
We first show the approximate capacity for the case of $M=2$ and independent channel states, then generalize this result for the case of any $M$ and independent channel states and, lastly, show
the approximate capacity for any $M$ and any correlation.

%
The CCDP is a special case of  the compound GP channel for which, unfortunately, not many results are available in the literature.
An achievable region for the  two-user compound GP channel is presented in \cite{nair2010achievability} where it is shown that
 using a common message potentially improves over extensions of the capacity achieving strategy for the GP channel in which the channel input is pre-coded
 against both channel states.
%
The CCDP was first proposed in \cite{LapidothCarbonCopying} where the authors consider both the binary and the Gaussian versions of the  $M$-user compound GP channel
and derive the first inner and outer bounds for these models.
We have previously considered the case of two users in \cite{RiniPhase14} and derived the approximate capacity for a certain set of correlations among circularly-symmetric Gaussian
 state sequences.
%
%
A model related to the CCDP channel is the state-dependent broadcast channel with a common message.
This model  is obtained from the CCDP channel by introducing two privates message to be communicated between the transmitter and each of the users. %
A first achievable region for this channel is obtained in  \cite{steinberg2005achievable} combining coding strategies for the GP channel and the broadcast channel \cite{MartonBroadcastChannel}.
Steinberg in \cite{steinberg2005coding} studies the channel in which one of the users is provided with the state sequence while the other user
observes a degraded channel output: capacity for this channel is obtained using bounding techniques inspired by the proof of the degraded broadcast channel capacity.
%

The remainder of the paper is organized as follows: in Sec. \ref{sec:Channel Model} we introduce the channel model, in Sec. \ref{sec:Related Results} we present the relevant results available in the literature.
In Sec. \ref{sec: 2WRDP} we derive the approximate capacity for the case $M=2$ and independent channel states
while in Sec. \ref{sec:The M-WRDP channel} we present the  approximate capacity for the case of any $M$ and independent channel states.
In Sec. \ref{sec:General CCDP channel} we present the approximate capacity of for the case of any pairwise correlation.
%
Sec. \ref{sec:Conclusion} concludes the paper.

\noindent
\underline{
Only sketches of the proofs are provided in the main text:}
\underline{the full proofs can be found in appendix.}

\section{Channel Model}
\label{sec:Channel Model}

\begin{figure}
\centering
\begin{tikzpicture}[node distance=2cm,auto,>=latex]
  \node at (-5,0) (source) {$W$};
  \node [int, right of = source,node distance = 1.4 cm](enc){Enc.};
  \node [sum, right of = enc,node distance = 1.7 cm](enc1){};
  \node [right of = enc1,node distance = 1cm](p11){+};
  \node [joint,right of = enc1,node distance = 1cm](p11){};
  \node [right of = p11,node distance = .75 cm](p12){+};
  \node [joint,right of = p11,node distance = .75 cm](p12){};
  \node [int, right of = p12,node distance = 1.8 cm](dec1){Dec. 2};
  \node [ right of = dec1,node distance = 1.5 cm](sink1){$\Wh_2$};
  \node [above of = p11,node distance = 1 cm](S1){$S_2^N$};
  \node [above of = p12,node distance = 1 cm](Z1){$Z_2^N$};
  \draw[->,line width=1pt] (S1) -- (p11) ;
  \draw[->,line width=1pt] (Z1) -- (p12) ;
  \draw[-,line width=1pt] (p11) -- (p12) ;
  \draw[->,line width=1pt] (p12) node[above, xshift =0.7 cm] {$Y_2^N$}-- (dec1) ;
 \node [above of = p11,node distance = 2 cm](p21){+};
 \node [joint,above of = p11,node distance = 2 cm](p21){};
  \node [right of = p21,node distance = .75 cm](p22){+};
  \node [joint,right of = p21,node distance = .75 cm](p22){};
  \node [int, right of = p22,node distance = 1.8 cm](dec2){Dec. 1};
  \node [right of = dec2,node distance = 1.5 cm](sink2){$\Wh_1$};
  \node [above of = p21,node distance = 1 cm](S2){$S_1^N$};
  \node [above of = p22,node distance = 1 cm](Z2){$Z_1^N$};
  \draw[->,line width=1pt] (S2) -- (p21) ;
  \draw[->,line width=1pt] (Z2) -- (p22) ;
  \draw[-,line width=1pt] (p21) -- (p22) ;
  \draw[->,line width=1pt] (p22) node[above, xshift =0.7 cm] {$Y_1^N$} -- (dec2) ;
  \node [below of = p11,node distance = 1 cm](p31){};
  \node [right of = p31,node distance = 1.5 cm](p32){$\vdots$};
  \node [below of = sink1,node distance = 1 cm](sink3){$\vdots$};
  \node [below of = dec1,node distance = 1 cm](dec3){$\vdots$};
  \node [below of = p31,node distance = 1.5 cm](p41){+};
  \node [joint,below of = p31,node distance = 1.5 cm](p41){};
  \node [right of = p41,node distance = .75 cm](p42){+};
  \node [joint,right of = p41,node distance = .75 cm](p42){};
  \node [int, right of = p42,node distance = 1.8 cm](dec4){Dec. M};
  \node [right of = dec4,node distance = 1.5 cm](sink4){$\Wh_M$};
 \draw[-,line width=1pt] (p41) -- (p42) ;
 \draw[->,line width=1pt] (p42) node[above, xshift =0.7 cm] {$Y_M^N$} -- (dec4) ;
 \node [above of = p41,node distance = 1 cm](SM){$S_M^N$};
 \node [above of = p42,node distance = 1 cm](ZM){$Z_M^N$};
 \draw[->,line width=1pt] (SM) -- (p41) ;
 \draw[->,line width=1pt] (ZM) -- (p42) ;
 %
 %
 \draw[-,line width=1pt] (enc) node[above, xshift =1.25 cm] {$X^N$}--(enc1) {};
 \draw[->,line width=1pt, bend left=90 ]  (enc1) |- (p21);
 \draw[->,line width=1pt] (enc1) -- (p11);
 \draw[->,line width=1pt] (enc1) |- (p41);
 \draw[->,line width=.5 pt] (source) -- (enc);
 \draw[->,line width=.5 pt] (dec1) -- (sink1);
 \draw[->,line width=.5 pt] (dec2) -- (sink2);
 \draw[->,line width=.5 pt] (dec4) -- (sink4);
 \draw[->,densely dotted,line width=.5 pt] (S1) -| (enc);
 \draw[->,densely dotted,line width=.5 pt] (S2) -| (enc);
 \draw[->,densely dotted,line width=.5 pt] (SM) -| (enc);
\end{tikzpicture}
\caption{The ``Carbon Copying on Dirty Paper'' (CCDP).}
\label{fig:CCDP channel}
\vspace{-.5 cm}
\end{figure}
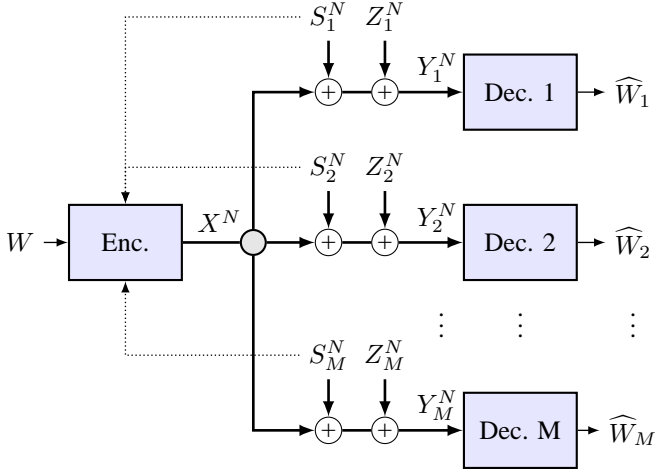

\label{def:M-user multicast channel Gelfand-Pisker with additive noise and additive state}
The $M$-user ``Carbon Copying on Dirty Paper'' (CCDP) channel,  also depicted in Fig. \ref{fig:CCDP channel},  is the compound GP channel
in which
the channel outputs are obtained as
\ea{
Y_m^N=X^N+c S_m^N+Z_m^N, \quad \quad m \in [1 \ldots M],
\label{eq:CCDP definition}
}
where $Z_m^N, \ \forall \ m$ is an iid Gaussian sequence with zero mean and unitary variance and $\{S_m^N, \ m \in [1 \ldots M] \}$ is an iid jointly Gaussian sequence with zero mean %
 and covariance matrix $\Sigma_S$ with
\ea{
1=\var[S_1] \leq \var[S_2] \ldots \leq \var[S_M],
\label{eq:state variance assumption}
}
where \eqref{eq:state variance assumption} is assumed without loss of generality.
The transmitter has anti-causal knowledge of $\{S_m^N, \ m \in [1 \ldots M] \}$ and is subject to the average power constraint $\sum_{n=1}^N \Ebb \lsb |X_n|^2 \rsb \leq N P$.

In the following we focus on the CCDP in which each state has unitary variance and each two states have the same correlation.
%
%
We term this model as ``Carbon Copying on Dirty Paper with Equivalent States'' (CCDP-ES), since  all the channel states are statistically equivalent.
%
The range of feasible values for the correlation $\rho$ is shown by the next lemma.

\begin{lem}
\label{lem:Feasible CCDP-ES}
Let the matrix $\Sigma_S$ be equal to
\ea{
\Sigma_S = (1-\rho) \Iv_{M,M} + \rho \ones_{M,M}    = \lsb \p{
1     &  \rho & \ldots  \\
\rho  &  1    &  \ddots        \\
\vdots & \ddots   &  \ddots      \\
} \rsb,
\label{eq:ES sigma matrix}
}
where $\Iv_{M,M}$ is the identity matrix of size $M$ and $\ones_{M,M}$  is the matrix of all ones of size $M \times M$,
then $\Sigma_S$ is positive defined for
\ea{
- 1/(M-1)  \leq \rho \leq 1.
\label{eq:positive defined covariance matrix}
}
\end{lem}
\begin{IEEEproof}
See App. \ref{app:Feasible CCDP-ES}.
\end{IEEEproof}
%
%
%
%
%
\begin{lem}
\label{lem:capacity decreasing scaling fading}
The capacity of the CCDP channel is decreasing in $c$.
\end{lem}
\begin{IEEEproof}
See App. \ref{app:capacity decreasing scaling fading}.
\end{IEEEproof}

This result is rather intuitive since capacity can only increase if we reduce the variance of the state.

\section{Related Results}
\label{sec:Related Results}
\medskip
\noindent
$\bullet$ {\bf  Carbon Copy onto Dirty Paper (CCDP)  channel.}
The channel model in \eqref{eq:CCDP definition} was originally introduced in \cite{LapidothCarbonCopying},
in which the authors derive a number of inner and outer bounds to capacity.
%

\begin{thm}{\bf Inner and outer bounds for the 2-CCDP channel with independent states \cite[Th. 3, Th. 4]{LapidothCarbonCopying}.}
\label{th:Inner and outer bounds 2-CCDP with independent states }
Consider the CCDP channel in \eqref{eq:CCDP definition} for $M=2$ and $\Sigma_S=\Iv_{2,2}$,  then capacity is upper bounded as
\ea{
\Ccal \leq R^{\rm OUT} = \lcb \p{
 \f 14 \log\lb\f{1+P}{ c^2/4+1} \rb  \\
\quad  +\f 1 4 \lb \f{1+P+c^2+2 c \sqrt{P}}{c^2/4 +1}\rb  & c^2 < 4 \\
 \f 14 \log(1+P)-\f 14 \log(c^2) \\
\quad  +\f 14 \log (1+P+c^2+2 c \sqrt{P} )  & c^2 \geq 4
} \rnone
\label{eq:outer lapidoth}
}
and lower  bounded as
\ea{
\Ccal \geq R^{\rm IN} = \lcb \p{
\f 12 \log \lb 1 + \f {P}{c^2/2+1}\rb   & c^2/2 \leq 1 \\
\f 12 \log \lb \f{P+c^2/2+1}{c^2}\rb \\
\quad + \f 14 \log \lb \f {c^2} 2 \rb & 1 \leq c^2/2 <P+1 \\
\f 14 \log (P+1) & c^2/2 \geq P+1
}
\rnone
\label{eq:inner lapidoth}
}

\end{thm}
A powerful bounding techniques is introduced in \cite{LapidothCarbonCopying} to derive the outer bound in \eqref{eq:outer lapidoth} while the inner bound in \eqref{eq:inner lapidoth} is obtained by having the transmitter pre-code against two linear combinations of the state sequences.

The outer bounding technique for the case of $M=2$ is also extended to the case of a general $M$.
\begin{thm}{\bf Outer bounds  for the M-CCDP channel with independent states \cite[Eq. (31)]{LapidothCarbonCopying}.}
Consider the CCDP channel in \eqref{eq:CCDP definition} for and $\Sigma_S=\Iv_{M,M}$, then capacity can be upper bounded as
\ea{
\Ccal
& \leq R^{\rm OUT} = \f 12 \log\lb P+c^2+2c \sqrt{P} \rb - \f{M-1}{2 M} \log c^2  \nonumber \\
&  \quad \quad -\f 1 {2 M}\log M - \lsb \f 1 {2M} \log \lb \f {c^2}{M(P+1)} \rb \rsb^+.
\label{eq:outer bound M independent states}
}
\end{thm}
Inner and outer bounds for the case $M=2$ are close for small values of $P$ but otherwise no capacity characterization is possible using the bounds in Th. \ref{th:Inner and outer bounds 2-CCDP with independent states }.
By generalizing the inner bound in \eqref{eq:inner lapidoth} to any $M$,  we can again show that inner and outer bound are close
only for small values of $P$.

\medskip
\noindent
$\bullet$ {\bf  Compound GP.}
The compound GP is a more general channel model than the CCDP:
%
in \cite{nair2010achievability} an attainable rate region for this model is obtained as:
\ea{
R^{\rm IN} \leq \max_{P_{X,V,U_1,U_2}} \min \lcb  I_1, I_2, \f 12 \lb I_1+I_2 - I(U_1;U_2| V, S_1,S_2) \rb  \rcb,
\label{eq:nair compound GP 1}
}
for $I_i = I(Y_i;U_i,V)-I(V,U_i;S_1,S_2), \ i\in\{1,2\}$. The variable  $V$ is a common message decoded at both receivers, while $U_1$ and $U_2$ are pre-coded against $S_1$ and $S_2$
respectively as in the GP channel.
%

\section{The 2-CCDP channel with independent, equal-variance states}
\label{sec: 2WRDP}
\begin{figure}
\centering
\begin{tikzpicture}
\node at (0,0) (w){};
\node [left of = w, distance=-2 cm ]{\includegraphics[width=0.52 \textwidth]{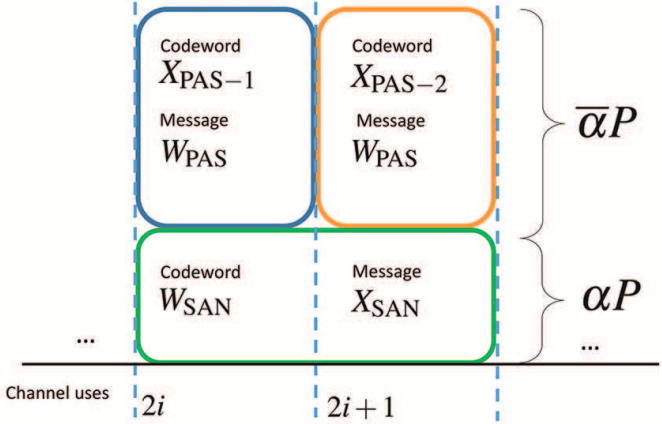}};
\end{tikzpicture}
\vspace{-0.5 cm}
\caption{A graphical representation of the capacity approaching  scheme in Th. \ref{th:Approximate capacity for the  2-CCDP with Gaussian independent states}.}
\vspace{-0.3 cm}
\label{fig:achievableSchemeSuperposition}
\end{figure}

We begin by deriving the approximate capacity for 2-CCDP-ES for $\rho=0$: this is  allows us to illustrate the main inner and outer bounding techniques while deferring more complex derivations to the latter sections.
%
%
In the derivation of the inner bound, we consider the same attainable strategy as in \cite{rini2014impact}, also depicted in Fig. \ref{fig:achievableSchemeSuperposition}:
the channel input is obtained as the superposition of three codewords:
(i) a bottom common codeword, $X_{\rm SAN}^N$ ($\rm SAN$ for \emph{State As Noise}) with power $\al P$, carries the message $W_{\rm SAN}$ with rate $R_{\rm SAN}$ and treats the state sequences $S_1^N$ and $S_2^N$ as additional noise while, and
(ii) two top private codewords, $X_{\rm PAS-1}^N$,$X_{\rm PAS-2}^N$ ($\rm PAS-i$ for \emph{Pre-coded Against State $S_i^N$}),
with power $\alb P$ for $\alb=1-\al$, pre-coded against $S_1^N$ and $S_2^N$ respectively and transmitted for half of the time each.
Since the $\var[S_1]=\var[S_2]$, the codeword $X_{\rm SAN}^N$ can be decoded at both receivers simultaneously.
On the other hand, $X_{\rm PAS-i}^N$ is decoded only at receiver $i$ since it is pre-coded  against the state $S_i^N$.
In order for the both decoders to decode the same amount of common information,  these codewords carry the same message $W_{\rm PAS}$ at rate $R_{\rm PAS}$.
As a result of these consideration, both receivers are able to correctly decode both $W_{\rm SAN}$ and $W_{\rm PAS}$.
%
%
%
%
thus attaining the transmission rate
\ea{
R^{\rm IN} = \f 12 \log \lb 1+ \f {\al P}{ c^2 +\alb P + 1}\rb  + \f 1 4 \log \lb 1+ \alb P \rb.
\label{eq:inner bound before power optimization}
}
The expression in \eqref{eq:inner bound before power optimization} can be maximized over $\al$, the ratio between the power of the common and the private codewords.
When $P+1\geq c^2$,  the optimal value of $\alb$ is  $(c^2-1)/P$, which corresponds to fixing the power of the private codewords to the same power
as the state sequence.
When $c^2>P+1$, instead, all the power is allocated to the private codewords and the scheme reduces to pre-coding for receiver~1 half of the time and pre-coding for receiver~2 the remaining portion of the time.

\begin{figure}
\begin{center}
\begin{tikzpicture}
\node at (0,0) {\includegraphics[trim=0cm 0cm 0cm 0cm,clip=true,scale=0.45]{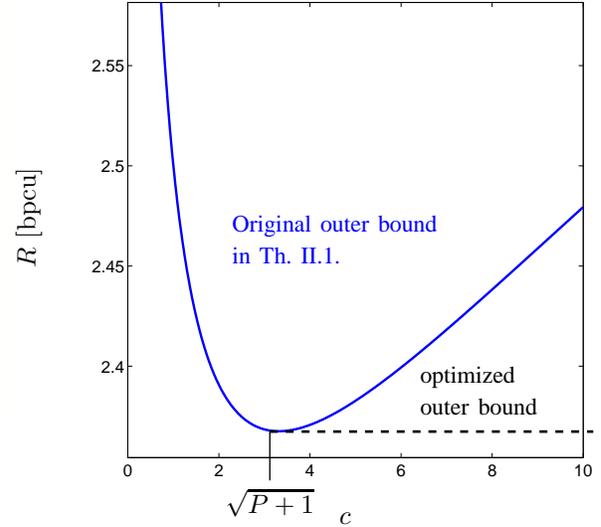}};
\vspace{-1 cm}
\node[rotate=90] at (-4.2,0.3) {{$R~ [\bpcu]$ }} ;
\node at (-1,-3.5) {$\sqrt{P+1}$};
\node at (0,-3.7) {$c$};
\draw[dashed,line width=1pt] (-1,-2.55) -- (+3.3,-2.55);
\draw[line width=.5 pt] (-1,-2.55) -- (-1,-3.2);
\node[rotate = 0, text width=3 cm] at (0,0) {\color{blue} \small Original outer bound in Th. \ref{th:Inner and outer bounds 2-CCDP with independent states }.};
\node[rotate = 0, text width=2 cm] at (2,-2) {\small optimized outer bound};
\vspace{-1 cm}
\end{tikzpicture}
\caption{ The outer bound in \eqref{eq:outer lapidoth} for $P=10$ and $c\in[0,10]$. }
\label{fig:outer bound}
\end{center}
\vspace{-.8 cm}
\end{figure}

With respect to  the outer bound, we are able to improve on the result of Th. \ref{th:Inner and outer bounds 2-CCDP with independent states } using the observation in Lem. \ref{lem:capacity decreasing scaling fading}: note that the outer bound  expression in \eqref{eq:outer lapidoth} for $c>4$ is not decreasing increasing in $c$, as shown Fig. \ref{fig:outer bound}.
For this reason it is possible to improve the outer bound by considering a channel with a parameter $c'=\min\{\sqrt{P+1},c\}\leq c$:
this channel has a larger capacity than the original channel but provides a tighter outer bound.
By comparing these inner and outer bound expressions, we can bound the capacity  to within  $1 \ \bpcu$.

\begin{thm}{\bf Approximate capacity for the  2-CCDP with independent, equal-variance states. \\}
\label{th:Approximate capacity for the  2-CCDP with Gaussian independent states}
Consider the 2-CCDP-IS channel in Fig. \ref{fig:CCDP channel} for $\Sigma_S=\Iv_{2,2}$, then an outer bound to capacity is
\ea{
& \Ccal \leq R^{\rm OUT}  = \nonumber \\
& \lcb \p{
 \f 1 2 \log(P+1)
 &  c^2 \leq 1 \\
 \f 12 \log(P+c^2+1) \\
\quad  - \f 14 \log(c^2+1)+\f1 2
& 1 < c^2 < P+1  \\
\f 1 4 \log(P+1)+1
& c^2 \geq P+1 \\
} \rnone
\label{eq: outer bound CCDP bernoulli}
}
and the exact capacity  $\Ccal$ is to within a gap of $1 \ \bpcu$ from the outer bound in \eqref{eq: outer bound CCDP bernoulli}.
\end{thm}
\begin{IEEEproof}
See. App. \ref{app:Approximate capacity for the  2-CCDP with Gaussian independent states}.
\end{IEEEproof}
The result in Th. \ref{th:Approximate capacity for the  2-CCDP with Gaussian independent states} is somewhat expected:
when the states in the 2-CCDP channel are independent, the best strategy is to send a common codeword at a power level larger than the channel state that can be decoded at both users
and a private codeword for each user, pre-coded against the state realization in the corresponding channel output.
In order for the private codeword to communicate the same message at the two receiver, this codeword must be time-shared between the
two receivers.
%
The major difficulty in proving theorem is therefore in deriving an outer bound which matches this intuitively optimal solution.
Before showing the approximate capacity of the CCDP-ES, we first show how to extend of the result it Thm. \ref{th:Approximate capacity for the  2-CCDP with Gaussian independent states} from the case of $M=2$ to the case of any number of users.

\section{The M-CCDP channel with independent, equal-variance states}
\label{sec:The M-WRDP channel}

\begin{figure}
\centering
\includegraphics[width=.52 \textwidth]{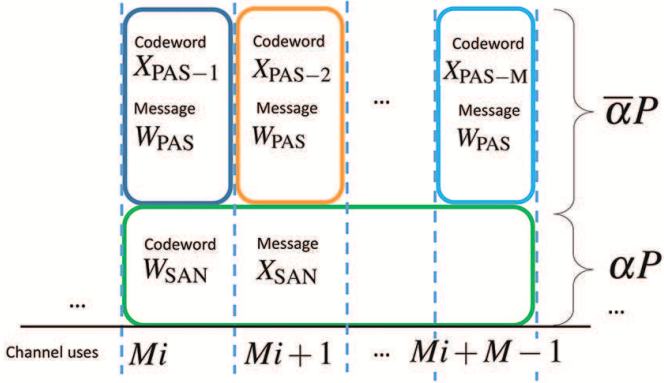}
\vspace{-1 cm}
\caption{A graphical representation of the capacity approaching scheme in Th. \ref{th:Approximate capacity for the  2-CCDP with Gaussian independent states}.}
\label{fig:achievableSchemeSuperpositionM}
\vspace{-0.8 cm}
\end{figure}

The approximate capacity of the M-CCDP channel with independent, equal-variance states is obtained through the appropriate extension of the inner and outer bounds in
Sec. \ref{sec: 2WRDP}.
A generalization of the inner bound in Fig. \ref{fig:achievableSchemeSuperposition} to the case of any number of users is rather straightforward:
we can modify the attainable strategy in Fig. \ref{fig:achievableSchemeSuperposition} as shown in Fig. \ref{fig:achievableSchemeSuperpositionM} and employ one common codeword
$X_{\rm SAN}^N$ at power $\al P$ and $M$ time-shared codewords $X_{\rm PAS-m}^N, \ m \in [1 \ldots M]$ of power $\alb P$, each pre-coded against the state sequence $S_m^N$.
All the codewords $X_{\rm PAS-m}^N$ convey the same message $W_{\rm PAS}$ and receiver $m$ decodes both the codeword  $X_{\rm SAN}^N$ and $X_{\rm PAS-m}^N$ so that,
at the end of the transmission, all the decoders can correctly decode both $W_{\rm SAN}$ and $W_{\rm PAS}$.
The rate that we can attain with this strategy is
\ea{
R^{\rm IN} = \f 12 \log \lb 1+ \f {\al P}{ c^2 +\alb P + 1}\rb  + \f 1 {2M} \log \lb 1+ \alb P \rb,
\label{eq:attainable rate WRDP with M}
}
which can again be maximized over $\al$.
In this case the optimal value of $\alb$ is
\ea{
\alb^*= \max \lcb 0,\min \lcb 1, \f{c^2+1-M}{P(M-1)} \rcb \rcb,
}
and the above scheme reduces to simple time-sharing and Costa pre-coding when $c^2>(M-1)(P+1)$.

The generalization of the outer bound in Th.  \ref{th:Approximate capacity for the  2-CCDP with Gaussian independent states} is rather more involved:
this can be accomplished by establishing a recursive bounding of the mutual information terms obtained from Fano's inequality and
using a very carefully-chosen genie side information for each decoder.
We refer the interested reader to \cite[App. D]{RiniISITstate-16} for the complete proof.
Again, the observation in Lem.  \ref{lem:capacity decreasing scaling fading} is employed to tighten the outer bound expression by optimizing over the state gain $c$.

\begin{thm}{\bf Approximate capacity M-user CCDP with independent, equal-variance states. \\}
\label{th:Approximate capacity M-CCDP with Gaussian independent states}
Consider the M-CCDP-IS channel in Fig. \ref{fig:CCDP channel} for $\Sigma_S=\Iv_{M,M}$
%
%
then an outer bound to capacity is
\ea{
& \Ccal \leq  R^{\rm OUT} =  \nonumber \\
& \lcb \p{
\f 12 \log \lb 1+\f  {P}{1+ c^2} \rb+\f 9 4&  M-1 \geq  c^2 \\
 \f 1 {2M} \log(1+P) &  \small {M-1 < c^2 \leq (M-1)(P+1)}   \\
 \quad +\f {M-1}{2 M } \log \lb {c^2} \rb+ \f 3 2  &  \\
 \f 1 {2M} \log(1+P)  +2   & c^2  > (M-1)(P+1)
} \rnone
\label{eq: outer bound CCDP M}
}
and the exact capacity  $\Ccal$ is to within a gap of $2.25 \ \bpcu$ from the outer bound in \eqref{eq: outer bound CCDP M}.
\end{thm}
\begin{IEEEproof}
See App. \ref{app:Approximate capacity M-CCDP with Gaussian independent states}.
\end{IEEEproof}

%
%

%
It is interesting to notice that pure time-sharing with no common codeword is approximatively optimal when $c^2  > (M-1)(P+1)$
that is when the state variance is roughly $M$ times stronger than the transmit power.
This occurs, intuitively, because the pre-log of the rate of the codeword $X_{\rm SAN}^N$ is $1/2$ while the pre-log of the codewords $X_{\rm PAS-m}$ is $1/2M$.

\section{The CCDP-ES channel}
\label{sec:General CCDP channel}
In this section we finally derive the approximate capacity of the CCDP-ES channel: the result relies, from a high-level viewpoint, on two observations:
(i) positive correlation among the states implies that there exists a common component which can be pre-coded against in the common codeword $X_{SAN}^N$, and
(ii) negative correlation among the states does not allow any improvement in the attainable rates with respect to the case of independent channel states.
To illustrate these points, note that the output of  the 2-CCDP-ES can be equivalently expressed as
\eas{
Y_1^N & = X + c \lb a  S_c^N + \sqrt{1-a} \St_1^N \rb + Z_1^N \\
Y_2^N & = X + c \lb \f{\rho} {a} S_c^N  + \sqrt{1-\f {\rho^2} {a^2} } \St_2^N \rb + Z_2^N,
}{\label{eq:common noise}}
for some $S_c,\St_1,\St_2 \sim  \Ncal(0,1), \ iid$, and any $a \in [-1,+1]$.
The choice $a=\sqrt{|\rho|}$ makes the term $S_c$ have the same scaling in both channel outputs: for the case of positive correlation this term can  be simultaneously
pre-coded at both receivers as in the WDP channel.
For of negative correlation, since the common term appears in with opposite sign in the two outputs, no coding advantage is possible.
%
%
\begin{thm}{\bf Approximate capacity for the general 2-CCDP-ES. \\}
\label{th:Approximate capacity for the  2-CCDP correlated}
Consider the general 2-CCDP channel with state covariance matrix $\Sigma_S$ as in \eqref{eq:ES sigma matrix} for $\rho$ satisfying \eqref{eq:positive defined covariance matrix},
 then capacity can be upper bounded as
\ea{
& \Ccal \leq  R^{\rm OUT}  = \\
& \lcb \p{
 \f 1 2 \log(P+1)
 &  c^2 \rhob^+ \leq 1 \\
 \f 12 \log(P+c^2 \rhob^++1) & 1 < c^2 \rhob^+ < P+1   \\
\quad  - \f 14 \log(\rhob^+ c^2)+\f1 2
\\
\f 1 4 \log(P+1)+\f 12
& c^2\rhob \geq P+1
} \rnone
\label{eq: outer bound 2-CCDP correlated}
}
for $\rhob^+=1-\max\{\rho,0\}$
and the exact capacity is to within $2.25 \ \bpcu$ from  the outer bound in \eqref{eq: outer bound 2-CCDP correlated}.
\end{thm}
\begin{IEEEproof}
See App. \ref{app:Approximate capacity for the  2-CCDP correlated}.
\end{IEEEproof}

The outer bound in \eqref{eq: outer bound 2-CCDP correlated} for $\rho>0$ is obtained by providing the common state $S_c$ as a side information to the receiver: the resulting channel
is then the same model as in Th. \ref{th:Approximate capacity for the  2-CCDP with Gaussian independent states} but with $c'=\rhob c$.
For the case of $\rho<0$, we rely on the fact that outer bound in Th. \ref{th:Approximate capacity for the  2-CCDP with Gaussian independent states}, when adapted to the case of correlated states, is increasing in the parameter $\rho$ and thus the case of $\rho=0$ provides a looser outer bound than the case of $\rho<0$.
%
%
The achievability proof for the case $\rho<0$ is the same as the achievability proof in Th. \ref{th:Approximate capacity for the  2-CCDP with Gaussian independent states},
 since this scheme  is not affected by correlation among the states.
For the case of $\rho>0$ we adapt the scheme in Th. \ref{th:Approximate capacity for the  2-CCDP with Gaussian independent states} by having the common codeword $X_{SAN}^N$
pre-coded against the common state sequence $c \sqrt{\rho} S_c^N$.

The decomposition of the channel outputs in \eqref{eq:common noise} in terms of a common component can be extended to the case of any users,
and the distinction between positive and negative pairwise correlation becomes clearer in this context.

For the case of positive correlation, a common term with variance $\rho$ can be extracted from all channel outputs by representing the channel states as
\ea{
S_m = \sqrt{\rho} S_c + \sqrt{1-\rho} \St_m \quad m \in [1 \ldots M],
\label{eq:M positive correlation}
}
for $S_c,\St_m \sim \Ncal(0,1), iid$.
As for the proof of Th. \ref{th:Approximate capacity for the  2-CCDP correlated}, the transmitter can simultaneously pre-code against the term $\sqrt{\rho} S_c$ at all the users as in
the WDP channel.

The case of negative correlation is more intriguing, since, in this case, the channel states can be represented as
\ea{
S_m= \sum_{j=m+1}^{N} \sqrt{\rho} \Sh_{mj} - \sum_{j=1}^{m-1}  \sqrt{\rho}\Sh_{jm} +  \sqrt{1-(N-1)\rho } \St_m,
\label{eq:negative rho case}
}
for $\Sh_{mj},\St_m \sim \Ncal(0,1) , [m,j]\in [1 \ldots M]^2, \ m>j$.
The representation in \eqref{eq:negative rho case} provides some intuition on the result in Lem. \ref{lem:Feasible CCDP-ES}: in order for the two states, $S_j$ and $S_k$ with $k>j$, to be
negatively correlated, they must share a term $\Sh_{jk}$ that does not appear in any other $S_m$.
This must be the case, otherwise this term would affect the correlation among $S_j,S_k$ and $S_m$.
Since each $S_m$ must be negatively correlated with other $N-1$ states, it must contain $N-1$  terms $\Sh_{mj}$ or $\Sh_{jm}$, each with variance $|\rho|$.
Given that the variance of $S_m$ is equal to one, we necessarily have that $|\rho|\leq 1/(N-1)$ or $\rho>-1/(N-1)$.
With the considerations in \eqref{eq:M positive correlation} and \eqref{eq:negative rho case} we can finally state the main result of the paper.

\begin{thm}{\bf Approximate capacity for the M-CCDP-ES. \\}
\label{th:Approximate capacity for the  M-CCDP with Gaussian independent states}
Consider the  M-CCDP channel in Fig. \ref{fig:CCDP channel} for $\Sigma_S$ as in \eqref{eq:ES sigma matrix}  for $\rho$ satisfying \eqref{eq:positive defined covariance matrix},
then capacity can be upper bounded as
\ea{
& \Ccal \leq  R^{\rm OUT}  = \nonumber \\
& \lcb \p{
\f 12 \log \lb 1+\f  {P}{1+ \rhob  c^2} \rb+\f 9 4    &  M-1 \geq  c^2 \rhob \\
 \f 1 {2M} \log(1+P) & {\small M-1 < c^2 \rhob  \leq (M-1)(P+1) }\\
 \quad +\f {M-1}{2 M } \log \lb {\rhob c^2} \rb+ \f 3 2  &  \\
 \f 1 {2M} \log(1+P)  +2   & \rhob c^2   > (M-1)(P+1)
} \rnone
\label{eq:outer M CCDP rho}
}
for $\rhob=1-\max\{0,\rho\}$  and the exact capacity is to within $2.25 \ \bpcu$ from  the outer bound in \eqref{eq:outer M CCDP rho}.
\end{thm}
\begin{IEEEproof}
{app:Approximate capacity for the  M-CCDP with Gaussian independent states}
\end{IEEEproof}

The difficulty in extending the result of Th. \ref{th:Approximate capacity for the  M-CCDP with Gaussian independent states} to the case of any correlation matrix $\Sigma_S$
lays  in the fact that, in this case, decoders have different decoding capabilities and therefore there are a number of ways in which the same set of public bits can be transmitted to
each receiver.
This can be accomplished by varying the time-sharing ratio for the private codeword for each receiver in the scheme in Fig. \ref{fig:achievableSchemeSuperpositionM}.
This optimization quickly becomes untractable and deriving a matching outer bound is challenging.

\section{Conclusion}
\label{sec:Conclusion}

In this paper we study the capacity of the carbon copying onto dirty paper  channel with equivalent states,
a variation of the classic dirty paper  channel in which the transmitted message is decoded at $M$ receivers,
each  observing  a linear combination of the input, Gaussian noise and one of $M$ possible state sequences.
These state sequences are  non-causally known at the transmitter and are statistically equivalent, being jointly Gaussian-distributed, with unitary variance and identical pairwise correlation.
%
Although inner and outer bounds to the capacity of this channel are available in the literature, no characterization of capacity was known.
We derive the capacity of this model to within 2.25 bits-per-channel-use for any channel and any pairwise correlation among the states.
In this model capacity can be approached with a rather simple strategy in which the input is composed of the superposition of two codewords:
a bottom, common codeword decoded at all users and a top, private codeword decoded at each receiver for a portion $1/M$ of the time  and pre-coded against the
channel state experienced at the given receiver.
%
The major contribution of the paper is in the derivation of an outer bound which closely approaches this intuitive inner bound.

Despite of our progress, the capacity of the channel in which the states have any jointly Gaussian  distribution remains unknown.
%
%
%
%
%
%
%
%

%
\bibliographystyle{IEEEtran}
\bibliography{steBib3,steBib1}

\newpage
\onecolumn
\appendices

\section{Proof of Lem. \ref{lem:Feasible CCDP-ES}}
\label{app:Feasible CCDP-ES}

For the matrix in \eqref{eq:ES sigma matrix}
the leading principal minor can be obtained through the matrix determinant lemma as
\ea{
\det((1-\rho)\Iv_{m,m}+\rho \ones_{1,m} \ones_{m,1} )=(1-\rho)^m \lb 1+ \f{m \rho} {1-\rho}\rb,
}
which is non-negative for
\ea{
\rho \geq -\f 1 {m-1}.
}
Consequently, all the leading principal minors of the matrix in \eqref{eq:ES sigma matrix}  are positive when
\ea{
\rho \geq \min_{m} \lcb -\f 1 {m-1} \rcb =-  \f 1 {M-1}.
\label{eq:princ minors}
}
Equation \eqref{eq:princ minors} together with the fact that  $\rho$ is necessarily bonded below one, we obtain the condition in \eqref{eq:positive defined covariance matrix}.

\section{Proof of Lem. \ref{lem:capacity decreasing scaling fading}.}
\label{app:capacity decreasing scaling fading}
Given the state sequence vector $\Sv^N=[S_1^N \ldots S_M^N]$, we can represent this sequence as being obtained as $\Sv_{1}^N+\Sv_2^N$ where
\ean{
\Sv_1^N & \sim \iid \Ncal(0, \rho \Sigma_S)\\
\Sv_2^N & \sim \iid \Ncal(0, \rhob \Sigma_S),
}
for $\Sv_1 \perp \Sv_2$ and  $\rhob=1-\rho$.

Consider now the channel in which the set sequences $\Sv_2^N$ is provided as a side information to the transmitter and all the receivers: the capacity of this channel must
necessarily be  larger than the capacity of the original channel, since this extra knowledge can be ignored.
The $m^{\rm th}$ receiver in the enhanced channel can produce the sequence $\Yt_{m}$ as
\ea{
\Yt_{m}^N
& =Y_{m}^N-c S_{2,m}^N \nonumber \\
& = X^N + c S_{1,m}^N + Z_m.
\label{eq:equivalent channel out}
}
The sequence $\Yt_{m}^N$ in \eqref{eq:equivalent channel out} is statistically equivalent to the channel in \eqref{eq:CCDP definition} for
\ea{
\ct=c \sqrt{\rho} \leq c,
}
where the state sequence $\Sv'$ is appropriately scaled so that \eqref{eq:state variance assumption} holds.
When considering the equivalent channel output $\Yt_{m}^N$, the sequences in $\So_2$  acts as a common information between the transmitter and the receivers and
thus does not increase capacity.
From these observations, we conclude that the capacity of the model with state gain $c$ and side information $\So_2$ is equivalent to the
capacity of the channel model in which the state gain is $\ct$.
This implies that the capacity increases as $c$ decreases and thus concludes the proof.
%

\section{Proof of Th. \ref{th:Approximate capacity for the  2-CCDP with Gaussian independent states}.}
\label{app:Approximate capacity for the  2-CCDP with Gaussian independent states}

A gap of $1 \ \bpcu$ for $P\leq 3$ or $c^2 \leq 3$ can be attained by either treating the state as noise or simple considering the trivial achievable point $R=0$,
so we consider here only the case $P>3$ and $c^2>3$.

The outer bound derivation initially follows steps similar to  that of \cite[Th. 3]{LapidothCarbonCopying} and is successively improved by using of the observation in
Lem. \ref{lem:capacity decreasing scaling fading}.
The inner bound is substantially the same inner bound as in \cite{rini2014impact} and relies on the superposition coding and binning: a base codeword treats the state
as noise and  and two top codewords which are each transmitted only for half of the time.
The first codeword  is  pre-coded against the channel state observed at one user while the second codeword is pre-coded against the state  observed at the second user.

\bigskip
\noindent
{\bf Capacity outer bound:}
\noindent
As in \cite[Th. 3]{LapidothCarbonCopying}, we have that the capacity of this channel can be upper bounded as
\eas{
N (R -\ep)
& \leq \min_{j \in \{1,2\}}I(Y_j^N;W) \\
& \leq \f 1 2 \lb H(Y_1^N)+H(Y_2^N) - H(Y_1^N|W)  - H(Y_2^N| W)\rb.
}
The sum of the positive entropy terms $H(Y_1^N)+H(Y_2^N)$ can be bounded as
\eas{
& H(Y_1^N)+H(Y_2^N)  \\
& \leq \f N 2 \log(P+c^2 + 2 c  \sqrt{P} +1) + \f N 2 \log(P+ c^2+ 2 c  \sqrt{P} +1)+ N  \log 2 \pi e
\label{eq:sum entropy bound 1} \\
& \leq N  \log 2 \pi e (P+c^2 +1)+ N +N  \log 2 \pi e,
\label{eq:sum entropy bound 2}
}{\label{eq:sum entropy bound}}
where \eqref{eq:sum entropy bound 1} follows from the Gaussian Maximizes Entropy (GME) property and \eqref{eq:sum entropy bound 2} follows from the fact that
\ea{
2(P+c^2)\geq  (\sqrt{P}+c)^2.
}
For the sum of negative entropy terms $-H(Y_1^N|W)  - H(Y_2^N| W)$ we have
\eas{
& -H(Y_1^N|W)  - H(Y_2^N| W) \\
& \leq -H(Y_1^N,Y_2^N| W) \\
& = -H(Y_2^N-Y_1^N,Y_2^N| W)
\label{eq:transform 1}\\
& = -H(c(S_1^N-S_2^N) +Z_2^N-Z_1^N, Y_2^N | W),
}{\label{eq:transform} }
where in \eqref{eq:transform 1} we have used the transformation
\ea{
\lsb
\p{Y_2^N-Y_1^N \\
Y_2^N
}
\rsb
=
\lsb \p{
-1 & 1  \\
0  & 1
}
\rsb
\cdot
\lsb
\p{
Y_1^N \\
Y_2^N
}
\rsb
\label{eq:jacobian 1}
}
which has jacobian equal one.
We now continue the series of inequalities in \eqref{eq:transform} as
\eas{
& = -H(c(S_2^N-S_1^N) +Z_2^N-Z_1^N| W) - H(Y_2^N| S_2^N-S_1^N +Z_1^N-Z_2^N, W) \\
& \leq -H(c(S_2^N-S_1^N) +Z_2^N-Z_1^N) - H(Y_2^N| S_1^N,S_2^N,W,Z_1^N-Z_2^N) \\
& \leq -H(c(S_2^N-S_1^N) +Z_2^N-Z_1^N) - H(Z_2^N| Z_1^N-Z_2^N).
\label{eq:bonuding difference states}
}
Since $Z_1^N \perp Z_2^N$, we obtain
\ean{
-H(Y_1^N|W)-H(Y_2^N| W) & \leq  N \lb -\f 12 \log 2 \pi e ( 2 c^2 +2) - \f 12 \log 2 \pi e \f 12 \rb  \\
&  = N \lb -\f 12 \log 2 \pi e ( c^2 +1) - \f 12 \log 2 \pi e  \rb.
}
The two above inequalities establish the outer bound
\ea{
R^{\rm OUT} & =\f12 \log \lb P+c^2+1\rb \nonumber \\
& \quad \quad  -\f 14 \log \lb c^2 +1 \rb +\f 1 2.
\label{eq:out independent}
}
Since the capacity of the  channel is decreasing in $c^2$, as shown in Lem. \ref{lem:capacity decreasing scaling fading}, we can optimize the outer bound in
\eqref{eq:out independent 1} over the set $c'\in[0,c]$.
In order to match the boundaries of the optimization in the inner and the outer bound, we choose to further loosen the outer bound in  \eqref{eq:out independent} to
\ea{
R^{\rm OUT} & =\f12 \log \lb P+c^2+1\rb \nonumber \\
& \quad \quad  -\f 14 \log \lb c^2 \rb +\f 12.
\label{eq:out independent 1}
}
The first derivative of  \eqref{eq:out independent 1} in $c^2$ is
\ea{
\f{\partial \  \eqref{eq:out independent 1} }{\partial c^2  } = - \f 1 4 \f{P+1-c^2}{(1+P+c^2) c^2},
}
which has a zero in $c^*=\sqrt{P+1}$. For $c^2=P+1$, the second derivation of \eqref{eq:out independent 1} in $c^2$ is positive: we can therefore set ${c'}=\min\{\sqrt{P+1},c\}$
and obtain a channel with a larger capacity but a tighter expression of the outer bound in \eqref{eq:out independent 1}.
The result of the optimization in $c$ correspond to bound in \eqref{eq: outer bound CCDP bernoulli}.
Note that, for the case  $c^2<1$ we use the trivial outer bound $\Ccal \leq \f 12 \log(P+1)$: since the variance of the state is $1$, the contribution of the state to the channel output
is minimal for this case.

\bigskip
\noindent
{\bf Capacity inner bound:}
\smallskip
Consider the transmission scheme in which the channel input $X^N$ is comprised of the superposition of the following codewords:

\noindent
$\bullet$ {\bf \ \!  (i)} the base codeword $X_{\rm SAN}^N$ ($SAN$ as in ``\emph{State As Noise}'') which treats the state as noise and

\noindent
$\bullet$ {\bf (ii)} the top codewords $X_{\rm PAS-i}^N$ ($PAS$ as in ``\emph{Pre-coded Against the State}'')  is pre-coded against the sequence $S_i^N$ for $i \in \{1,2\}$.
%
Additionally $X_{\rm PAS-1}^N$ is transmitted for the first half of the time, while $X_{\rm PAS-2}^N$ is transmitted for the second half of the time.
The codewords $X_{\rm PAS-i}^N$ are superimposed over the codeword $X_{\rm SAN}^N$: receiver $i$ jointly decodes $X_{\rm SAN}^N$ and $X_{\rm PAS-i}^N$.
%
All the codewords are iid Gaussian-distributed: $X_{\rm SAN}^N$ has power $\al P$ while $X_{\rm PAS-i}$ have both power $\alb P$ for some $\al \in [0,1], \ \alb=1-\al$.
The common codeword $X^{\rm SAN}$ attain the rate
\ea{
R_{\rm SAN}
& = I(X_{\rm SAN}; Y_i)  \nonumber \\
& = \f 12 \log \lb 1+\f {\al P}{1 + c^2 + \alb P}  \rb
}
and is used to communicate the messages $W_{\rm SAN} \in [1 \ldots 2^{N R_{\rm SAN}}]$ to both  users simultaneously.
The two private codewords $X_{\rm PAS-1}$ and $X_{\rm PAS-2}$  each attain the rate
\ea{
R_{\rm PAS} & = I(Y_1; U_1|X_{\rm SAN}) - I(U_1;S_1) \nonumber \\
            & = I(Y_2; U_2|X_{\rm SAN}) - I(U_2;S_2)
}
and encode the same message $W_{PAS} \in [1 \ldots 2^{N R_{\rm PAS}}]$. Note that the message $W_{PAS}$ is sent twice, since it is reliably communicated to the first users in the first
half  of the transmission and to the second user the second half of the transmission.
%
Combining the rate of the common and the private message, we conclude that the overall attainable rate is
\ea{
R^{\rm IN}(\al) =
&   \f 12 \log \lb 1 + \f{\al P} {1 +\alb P+ c^2} \rb + \f 14 \log \lb 1 + \alb P  \rb
\label{eq:Interference as Noise + Binning Inner Bound},
}
for any $\al \in [0,1]$.
The optimization over $\al$ yields that the optimal value
\ea{
\alb*=\lcb \p{
0               & c^2 < 1 \\
\f {c^2-1} P    & 1 \leq c^2 < P+1\\
1               & c^2 \geq P+1
}\rnone
}
and the corresponding optimal rates
\ea{
R^{\rm IN} = \lcb \p{
\f 12 \log \lb 1+ \f P {c^2 +1}\rb               & c^2 < 1 \\
\f 12 \log \lb 1+ c^2 + P\rb -\f 14 \log(c^2)- \f 12   & 1 \leq c^2 < P+1\\
\f 14 \log (P+1)               & c^2 \geq P+1
}\rnone
\label{eq:overall rate app}
}

\bigskip
\noindent
{\bf Gap between inner and outer bound:}
\smallskip
For the case $c^2 \leq 1$ we notice that the distance between inner and outer bound is at $1/2 \ \bpcu$ using simple considerations on the shape of the capacity region.
For  the remaining cases, inner and outer bounds can be compared directly: the gap is $1 \ \bpcu$  for $c^2\geq P+1$ and also $1 \ \bpcu$  for the case  $1 \leq c^2 < P+1$.
We therefore conclude that, regardless of the channel parameters,  the outer bound can be attained to within $1 \ \bpcu$.

\section{Proof of Th. \ref{th:Approximate capacity M-CCDP with Gaussian independent states}.}
\label{app:Approximate capacity M-CCDP with Gaussian independent states}

The proof is  an extension of the proof of Th. \ref{th:Approximate capacity for the  2-CCDP with Gaussian independent states} and thus relies on similar inner and outer
bounding techniques.
The first part of the outer bound derivation follows the derivation of \cite[Eq. (31)]{LapidothCarbonCopying} but later we employ a recursive bounding of the mutual information terms
to come to a tighter bounding.
%
On the other hand the inner bound is  a rather straight forward extension of the bound in Th. \ref{th:Approximate capacity for the  2-CCDP with Gaussian independent states} in which
a bottom codeword and multiple top, pre-coded codewords are used to communicate the common message.

As for the proof in App. \ref{app:Approximate capacity for the  2-CCDP with Gaussian independent states}, we only need to consider the case $P>3$ and $c^2>3$ since the capacity region
is smaller than $1 \bpcu$ otherwise.

\bigskip
\noindent
{\bf Capacity outer bound:}
\smallskip
As in \cite[App. 3.C]{LapidothCarbonCopying}, we write
\eas{
N(R-\ep)
& \leq \min_{m \in [1 \ldots M]}  I(Y_m^N;W)\\
& \leq \f 1 M \sum_{m=1}^M I(Y_m^N;W) \\
& \leq  \max_{m \in [1 \ldots M]} H(Y_m^N) - \f 1 M \sum_{i=1}^M  H(Y^N|W)  \\
& \leq  \f N 2 \log (P+c^2+2 c \sqrt{P}+1)+\f N 2 \log (2 \pi e) - \f 1 M  \sum_{m=1}^M  H(Y_m^N|W).
\label{eq:label M case}
}

We now proceed in establishing a recursion by defining the term $T_m$ as
\ea{
T_m \triangleq \sum_{i=m}^M  H(Y_m^N|W),
\label{eq:def Tm}
}
which allows us to rewrite \eqref{eq:label M case} as
\ea{
N(R-\ep) \leq \f N 2 \log (P+c^2+2 c \sqrt{P}+1)+\f  N2 \log(2 \pi e) -T_1.
\label{eq:label M case 2}
}
The term $T_1$ can now be rewritten as
\ea{
- T_1
& = -H(Y_1^N|W)  - H(Y_2^N| W) -T_3.
}
We have seen in  \eqref{eq:transform} that  the difference $-H(Y_1^N|W)  - H(Y_2^N| W)$ can be bounded as follows:
\ea{
- T_1& \leq  -H(c(S_1^N-S_2^N) +Z_2^N-Z_1^N, Y_2^N | W)-T_3.
}
%
%
Since the noise terms are $Z_i$ to be indented and identically distributed we have:
\eas{
- T_1 & = -H(c(S_1^N-S_2^N)) - H(Y_2^N | S_1^N-S_2^N,W)- H(Y_3^N| W)- T_4 \\
& = - \f N 2 \log(2 c^2) - H(Y_2^N | S_1^N-S_2^N,W)- H(Y_3^N| W)- T_4 \\
& \leq - \f N 2 \log(2 c^2) - H(Y_2^N, Y_3^N| S_1^N-S_2^N,W)- T_4
\label{eq:transformation 2}\\
& \leq - \f N 2 \log(2 c^2) - H(Y_3^N-Y_2^N, Y_3^N|S_1^N-S_2^N,W)- T_4 \\
& \leq - \f N 2 \log(2 c^2) - H(c(S_3^N-S_2^N), Y_3^N| S_1^N-S_2^N,W)- T_4 \\
& = - \f N 2 \log(2 c^2) - H(c (S_3^N-S_2^N) | S_1^N-S_2^N) -H(Y_3^N| S_1^N-S_2^N,S_3^N-S_2^N,W)- T_4 \\
& \leq- \f N 2 \log(2 c^2) - \f N 2  \log \lb \f 3 2 c^2 \rb-H(Y_3^N| S_1^N-S_2^N,S_3^N-S_2^N,W)- T_4.
}{\label{eq:recursion M CCDP}}
A recursion can now be established on the same lines as \eqref{eq:recursion M CCDP} to bound all the terms in the summation $T_1$:
let $\Delta^N_1=\zerov_{1,N}$ and define $\Delta^N_i, \ i>1$ as
\ea{
\Delta^N_i \triangleq S_i^N-S_{i-1}^N,
}
then we can write
\eas{
T_1 & \leq   \sum_{i=2}^{K} H(c \Delta^N_i| \Delta^N_1 \ldots \Delta^N_{i-1}) - H(Y_{K}^N|\Delta^N_1 \ldots \Delta^N_K) - T_{K+1} \\
& =  \sum_{i=2}^{K} H(c \Delta^N_i| \Delta^N_1 \ldots \Delta^N_{i-1}) - H(Y_{K}|\Delta^N_1 \ldots \Delta^N_K,W)- H(Y_{K+1}|W) - T_{K+2} \\
& \leq  \sum_{i=2}^{K} H(c \Delta^N_i| \Delta^N_1 \ldots \Delta^N_{i-1}) - H(Y_{K}, Y_{K+1}|\Delta^N_1 \ldots \Delta^N_K,W) - T_{K+2}.
}
 By proceeding in this manner up to $K=M$ we come to the bound
\eas{
- T_1
& \leq \sum_{i=2}^{M} -H(c \Delta^N_i| \Delta^N_1 \ldots \Delta^N_{i-1}) - H(Y_M^N|\Delta^N_1 \ldots \Delta^N_{M},W) \\
& \leq \sum_{i=2}^{M} -H(c \Delta^N_i| \Delta^N_1 \ldots \Delta^N_{i-1}) - H(Z_M^N) \\
& \leq \sum_{i=2}^{M} -H(c \Delta^N_i| \Delta^N_1 \ldots \Delta^N_{i-1}) - \f N 2 \log(2 \pi e).
}
We are now left to evaluate the intermediate terms in the summation:
\ea{
H(c \Delta^N_i| \Delta^N_1 \ldots \Delta^N_{i-1})=  \f 1 2 \log(c^2) + H(\Delta^N_i| \Delta^N_1 \ldots \Delta^N_{i-1}).
\label{eq:recursion delta}
}
The correlation matrix  among the entries of the vector $[\Delta^N_2 \ldots \Delta^N_M]$ is
\ea{
\Sigma_{\Delta^N} =  \lsb\p{
2       & -1        &  0 & 0    &  &  \ldots & 0 \\
-1      &  2        & -1 &     & & \ldots & 0 \\
0       & -1        &  2 & - 1 & & \ldots & 0 \\
\vdots  &           & \ddots   & \ddots  & \ddots  & & \vdots  \\
 &&0 & -1 &2 & -1 & 0\\
\vdots   &&&0 & -1 &2 & -1\\
0       & \ldots  &&&0 & -1 &2 \\
}
\rsb
\label{eq:matrix 2 -1 0}
}
and thus we conclude that
\eas{
 - H( \Delta^N_i| \Delta^N_1 \ldots \Delta^N_{i-1})
 & =  -\f N2 \log \lb 2 -   \lsb - 1 \ldots -1 \rsb   \cdot \lsb\p{2 & -1 &  0\\ -1 & \ddots & \ddots \\ 0 &   \ddots  & } \rsb \cdot \lsb \p{ -1 \\ \vdots \\ -1 \\ } \rsb  \rb  \\
& =- \f N2 \log \lb 2  -   \f {i-1} {i} \rb
\label{eq:band matrix 2}\\
&  \leq  - \f N2 \log 1 =0,
}{\label{eq:band matrix}}
where \eqref{eq:band matrix 2} follows from the properties of symmetric tri-diagonal matrices.

With the bounding in  \eqref{eq:band matrix}, we obtain the outer bound
\ea{
R^{\rm OUT}
& \leq \f 1 2 \log(1+P+c^2) - \f {M-1} {2M} \log c^2 + \f 12 \log 2 \pi e \lb 1-\f 1 {2M} \rb \nonumber \\
& \leq \f 1 2 \log(1+P+c^2) - \f {M-1} {2M} \log c^2 + \f 3 2,
\label{eq:outer M CCDP}
}
and, as for the proof of  Th. \ref{th:Approximate capacity for the  2-CCDP with Gaussian independent states}, this outer bound can be optimized over $c$ in the interval $c' \in[0,c]$.
The derivative of \eqref{eq:outer M CCDP} in $c^2$ is equal to zero in
\ea{
c^*= \sqrt{(M-1)(P+1)},
}
while the second derivative is positive in this point.
Having that ${c^2}^*= (M-1)(P+1)$ is a minimum of the outer bound in \eqref{eq:outer M CCDP} when $c^2 \geq (M-1)(P+1)$,
we obtain the outer bound expression in \eqref{eq:outer M CCDP} for $c^2> M-1$.

For the interval $M-1 \geq c^2$ we bound the expression in \eqref{eq:outer M CCDP} as follows:
\eas{
& \f 1 2 \log(1+P+c^2) - \f {M-1} {2M} \log c^2 + \f 3 2   \nonumber \\
& \leq  \f 1 2 \log \lb 1 +\f {P}{1+c^2} \rb + \f 1 2 \log (1+c^2) - \f {M-1} {2M} \log c^2 +\f 3 2 \\
& \leq  \f 1 2 \log \lb 1 +\f {P}{1+c^2} \rb + \f 1 2 \log (2c^2) - \f {M-1} {2M} \log c^2 + \f 3 2
\label{eq:bound CCDP 1}\\
&  = \f 1 2 \log \lb 1 +\f {P}{1+c^2} \rb +\f 1 {2M} \log (c^2) + 2 \\
&  = \f 1 2 \log \lb 1 +\f {P}{1+c^2} \rb + \f 1 {2M} \log (M-1) + 2\\
&  \leq \f 1 2 \log \lb 1 +\f {P}{1+c^2} \rb + \f 1 4  + 2,
\label{eq:bound CCDP 2}
}
where \eqref{eq:bound CCDP 1} follows from the assumption that $c^2>1$ and \eqref{eq:bound CCDP 2} from the fact that $x^{-1} \log (x-1)$ has a maximum in $x=4$
when $x$ is integer valued.
Combining these results, we obtain the desired outer bound in \eqref{eq: outer bound CCDP bernoulli}.

\bigskip
\noindent
{\bf Capacity inner bound:}
\smallskip
Consider an inner bound which extends the inner  Th. \ref{th:Approximate capacity for the  2-CCDP with Gaussian independent states}  and in which the inner bound is composed
of the superposition of two codewords:

\noindent
$\bullet$ {\bf \ \!  (i)} the base codeword $X_{\rm SAN}^N$ ($SAN$ as in ``\emph{State As Noise}'') which treats the state as noise and

\noindent
$\bullet$ {\bf (ii)} the top codewords $X_{\rm PAS-i}^N$ ($PAS$ as in ``\emph{Pre-coded Against the State}'')  is pre-coded against the sequence $S_i^N$ for $i \in \{1 \ldots M\}$,
each  transmitted only for a portion $1/M$ of the time.
%
%
The rate achieved by each user with this scheme is
\ea{
R^{\rm IN} =
& \max_{ \al \in [0,1]}  \f 12 \log \lb 1 + \f{\al P} {1 +\alb P+ c^2} \rb + \f 1{2M} \log \lb 1 + \alb P  \rb
\label{eq:Interference as Noise + Binning Inner Bound 2}.
}
%
The optimization over $\al$ yields the achievable rate
\ea{
R^{\rm IN} = \lcb \p{
 \f 12 \log \lb 1 + \f{P} {1+c^2} \rb & M-1 >  c^2 \\
\f 12 \log (P+c^2  +1)   &  {M-1}  \leq  c^2  \leq (M-1) (P+1) \\
\quad - \f {M-1}{2 M } \log \lb c^2 \rb-\f 12  \\
 \f {1} {2M} \log(1+P)     &
}\rnone
\label{eq:inner bound strong}
}

\bigskip
\noindent
{\bf Gap between inner and outer bound:}
\smallskip
Consider the case where $M>2$, and compare the expression in  \eqref{eq:outer M CCDP} and in \eqref{eq:inner bound strong}.
The gap for $M-1>c^2$ is at most $2 \ \bpcu$, for the case ${M-1}  \leq  c^2  \leq (M-1) (P+1)$  and it is $2 \ \bpcu$  also when $c^2  > (M-1)(P+1)$ is $1/2M$.
The largest gap between inner and outer bound is $2.25 \ \bpcu$ and is attained for $M-1 \leq c^2$.

\section{Proof of Th. \ref{th:Approximate capacity for the  2-CCDP correlated}}
\label{app:Approximate capacity for the  2-CCDP correlated}

Lets consider the case of positive and negative correlation separately, since they require a separate derivation

\bigskip
\noindent
{\bf  Approximate capacity for $\rho>0$:}
\smallskip
For the outer bound we simply consider the outer bound in \eqref{eq: outer bound CCDP bernoulli} obtained by providing $S_c$ to both decoders: after this term
 is stripped from the channel output, the receivers  obtain the same model as in  Th. \ref{th:Approximate capacity for the  2-CCDP with Gaussian independent states} but with a state with smaller variance, that is $1-\rho$ instead of $1$.
By absorbing this factor in $c$, we obtain the outer bound in \eqref{eq:outer M CCDP rho}.

For the inner bound consider the generalization of the inner bound in App. \ref{app:Approximate capacity for the  2-CCDP with Gaussian independent states} in which the base codeword $X_{SAN}^N$ is pre-coded  against the state $S_c^N$ so that the rate is can be transmitted at rate
\ea{
R^{\rm SAN} = \f 12 \log \lb 1 +  \f {\al P} {1 + \alb P + (1-\rho)Q }\rb.
}
With adjustment to the attainable scheme in Th. \ref{th:Approximate capacity for the  2-CCDP with Gaussian independent states}, we see that the region in \eqref{eq: outer bound 2-CCDP correlated} can be attained to within $1 \ \bpcu$.

\bigskip
\noindent
{\bf  Approximate capacity for $\rho<0$.}
\smallskip

Note that the correlation affects the derivation of the outer bound in App. \ref{app:Approximate capacity for the  2-CCDP with Gaussian independent states} only in the derivation of the term \eqref{eq:bonuding difference states} where in can be noted that  $H(c(S_2^N-S_1^N) +Z_2^N-Z_1^N)$ is decreasing  in the correlation $\rho$
\ea{
H(c(S_2^N-S_1^N) +Z_2^N-Z_1^N) = \f 12 \log \lb 2 c^2(1-\rho)   \rb,
}
accordingly we have that outer bound for independent states in an outer bound for the case of negatively correlated states.
The negative correlation, also, does not affect the inner bound in Th. \ref{th:Approximate capacity for the  2-CCDP with Gaussian independent states} so that  the same rate as in
\eqref{eq:overall rate app} is attainable.
With these two considerations we see that the capacity for the case of negative correlation can be approached in the same manner as the case of independent states.

\section{Proof of Th. \ref{th:Approximate capacity for the  M-CCDP with Gaussian independent states}}
\label{app:Approximate capacity for the  M-CCDP with Gaussian independent states}
The case of positive correlation straightforwardly extends from the proof of Th. \ref{th:Approximate capacity for the  2-CCDP correlated} in App. \ref{app:Approximate capacity for the  2-CCDP correlated}.

For the case of negative correlation, we shall show that the recursion in \eqref{eq:recursion delta} is not affected by the negative correlation and that the
value of the entropy term in \eqref{eq:band matrix} are decreasing in the value of the correlation.

Note that the covariance matrix in \eqref{eq:matrix 2 -1 0} is not affected by the correlation, since
\ean{
\var \lsb \Delta_i \rsb
& = \var \lsb S_i-S_{i-1} \rsb \\
& = 2 (1-\rho),
%
}
and equivalently
\ean{
\Ebb \lsb \Delta_i \Delta_{i+1}\rsb
& =\Ebb \lsb \lb S_i-S_{i-1}\rb \lb S_{i+1}-S_i \rb \rsb  \\
& =\rho-1-\rho+\rho \\
& =-(1-\rho),
}
while for $j>i+1$
\ean{
\Ebb \lsb \Delta_i \Delta_{j}\rsb
& =\Ebb \lsb \lb S_i-S_{i-1}\rb \lb S_{j}-S_{j-1} \rb \rsb  \\
& =\rho -\rho-\rho+\rho \\
& =0,
}
so that
\ea{
 - H( \Delta^N_i| \Delta^N_1 \ldots \Delta^N_{i-1})
& =- \f N2 \log \lb (1-\rho)^N \lb 2  -   \f {i-1} {i}\rb \rb
\label{eq:band matrix corr} \\
&  \leq  - \f N2 \log 1 =0, \nonumber
}
%
where the expression in \eqref{eq:band matrix corr} is increasing in $\rho$ and thus  once again obtain that the outer bound for negative correlation is upper bounded
 by the outer bound for independent states.
As in the proof of Th. \ref{th:Approximate capacity for the  2-CCDP correlated} in App. \ref{app:Approximate capacity for the  2-CCDP correlated}, the inner bound is not affected by negative correlation: we  therefore conclude that the capacity for the case of negative correlated states is bounded in the same manner as in Th. \ref{th:Approximate capacity for the  M-CCDP with Gaussian independent states}.

\end{document}